\documentclass[nocopyright]{cimento}

\usepackage{graphicx}
\usepackage{amsmath}
\usepackage{amssymb}
\usepackage{latexsym}
\usepackage{tabularx} % per tabelle
\usepackage{comment}

\usepackage[dvipsnames]{xcolor}

\title{A continuity-equation-compatible basis for Effective Field Theory of Dark Energy}
\author{Federico Armato\from{ins:x}\from{ins:y}\thanks{federico.armato@ge.infn.it}, Edoardo Martinelli\from{ins:w}\from{ins:z}, Marco Raveri\from{ins:x}\from{ins:y}}
\instlist{
\inst{ins:x} Department of Physics, University of Genoa, Via Dodecaneso 33, 16146 Genoa, Italy
\inst{ins:y} INFN, Genoa Section, Via Dodecaneso 33, 16146 Genoa, Italy
\inst{ins:w} Department of Physics and Astronomy, University of Bologna, Via Gobetti 93/2, 40129 Bologna, Italy
\inst{ins:z}INAF, Astrophysics and Space Science Observatory Bologna, Via Gobetti 93/3, 40129 Bologna, Italy}



\begin{document}

\maketitle

\begin{abstract}
The Effective Field Theory (EFT) of Dark Energy (DE) is a model-independent framework that allows for the description of a wide class of dark energy and modified gravity models. This is achieved by extending the Hilbert–Einstein action through the introduction of time-dependent functions. However, the choice of these functions — and thus the operator basis — is not unique. In this paper, we propose a physically motivated constraint based on the continuity equation, leading to a \emph{continuity-equation-compatible} (CEC) basis that ensures a clear physical interpretation of the background EFT operators.
\end{abstract}

%%%%%%%%%%%%%%%%%%%%%%%%%%%%%%%%%%%%%%%%%%%%
\section{Introduction}
%%%%%%%%%%%%%%%%%%%%%%%%%%%%%%%%%%%%%%%%%%%%

For decades, the late-time acceleration of our Universe has been a fascinating subject of study for physicists and cosmologists~\cite{Adam_1998,Perlmutter_1999}. This has driven the development of alternative theories of gravity in an effort to better understand the physical laws underlying cosmic acceleration~\cite{Silvestri_2009}.\\
In recent years, a powerful framework has emerged that promises to address this problem: the Effective Field Theory of Dark Energy (EFT of DE)~\cite{Piazza:2013xia,Bloomfield_2013}.
This formalism is capable of capturing the dynamics of a broad class of modified gravity theories, enabling their classification based on the imprints they leave on both the smooth background expansion of the Universe and the evolution of linear perturbations.\\
While EFT of DE has proven to be a valuable tool in studying late-time acceleration, several challenges remain. 
One of the key difficulties lies in identifying a suitable parametrization of the functions that define the theory. 
In this work, we address this issue by introducing a parametrization guided by physical constraints - specifically, by selecting a basis for background operators that satisfies a continuity equation, in presence of a running of the Planck mass. 
We call this as a \emph{continuity-equation-compatible} (CEC) basis.

%%%%%%%%%%%%%%%%%%%%%%%%%%%%%%%%%%%%%%%%%%%%
\section{The Effective Field Theory of Dark Energy} \label{sec:eft.intro}
%%%%%%%%%%%%%%%%%%%%%%%%%%%%%%%%%%%%%%%%%%%%

The Effective Field Theory of Dark Energy is a model-independent framework that encompasses a broad class of single-field dark energy and modified gravity models, providing a unified description of their low-energy dynamics~\cite{Piazza:2013xia,Bloomfield_2013}.\\
In the EFT framework, modifications to the standard $\Lambda$CDM model at both the background and perturbation levels are described in terms of geometric objects defined in the unitary gauge, the reference frame where scalar field fluctuations vanish. As a result, the EFT action includes terms that explicitly depend on time due to the breaking of time diffeomorphism invariance.\\
To describe the dynamics of the universe at both the background and linear perturbation levels, it is sufficient to truncate the EFT action at quadratic order in perturbations:
\begin{align} \label{eq:eft.total.action}
S_{\rm tot} = S^{(0)}_{\rm EFT} + S^{(2)}_{\rm EFT} + S_{\rm m}[g_{\mu\nu}, \psi_{\rm m}]
\end{align} 
where $S^{(0)}_{\rm EFT}$ and $S^{(2)}_{\rm EFT}$ denote the part of the EFT action responsible for the evolution of the background and linear perturbations respectively.
$S_{\rm m}$ denotes the matter action for all relevant matter fields, $\psi_{\rm m}$.\\
The background action, in the notation of~\cite{Frusciante:2014xia}, reads:
\begin{align} \label{eq:eft.background.action}
S^{(0)}_{\rm EFT} =& \frac{1}{2}\int d^{4}x \sqrt{-g} \left[ M_{p}^{2}\Omega(t) R - 2c(t) \delta g^{00} +2 \Lambda(t) \right]
\end{align}
where $M_{p}$ is the bare Planck mass, $R$ is the Ricci scalar, computed from the metric $g_{\mu\nu}$ to which all matter species are coupled, and $\Omega (t)$, $c(t)$ and $\Lambda (t)$ are time-dependent functions known as background EFT functions, which characterize deviations from the standard model. The canonical Hilbert-Einstein action, and consequently the $\Lambda CDM$ model, can be obtained by setting $\Lambda(t)=\Lambda$, $c(t)=0$ and $\Omega(t)=1$.\\
Starting from the background EFT action in Eq.~\eqref{eq:eft.background.action}, one can derive the corresponding equations of motion. These can then be evaluated on a homogeneous and isotropic spacetime by using the Friedmann–Robertson–Walker (FRW) metric and the energy–momentum tensor of a perfect fluid, yielding the modified Friedmann equations:
\begin{align} \label{eq:friedmann.equations}
    3 M_{p}^{2} \Omega \left[ H^{2}+\frac{K}{a^{2}}+H\frac{\dot{\Omega}}{\Omega} \right]&=2c-\Lambda + \rho_{\rm m} \nonumber\\
     M_{p}^{2}\Omega \left[ 2\dot{H}+3H^{2}+\frac{K}{a^{2}}+\frac{\ddot{\Omega}}{\Omega}+2H\frac{\dot{\Omega}}{\Omega} \right] &= -\Lambda - P_{\rm m}
\end{align}
where $t$ denotes cosmic time, overdots represent derivatives with respect to $t$,  $H=\dot{a}/a$ is the Hubble rate, $K$ the spatial curvature, and $\rho_{\rm m}$ and $P_{\rm m}$ stand for the total matter density and pressure, respectively. Note that all time dependencies have been omitted and will be left implicit in what follows.\\
As we can see, in the EFT framework at the background level we have four free functions of time $\Omega(t)$, $c(t)$, $\Lambda(t)$ and $H(t)$. As usual we have two constraint equations, the modified Friedmann equations.  These constraints are differential algebraic equations, which typically means that, modulo boundary conditions, we have the freedom to specify two of these four functions and get the other two from the constraints.\\
The choice of which four functions to use is arbitrary  and we refer to different choices as EFT functions bases. Since we have this freedom all different bases are clearly physically equivalent. However in one basis or the other physical meaning of different functions might vary and can be more or less obscure.

%%%%%%%%%%%%%%%%%%%%%%%%%%%%%%%%%%%%%%%%%%%%

%\section{Background parametrizations} \label{sec:literature}
\section{A continuity-equation-compatible basis} \label{sec:optimal.parametrization}
In the EFT framework, a common approach to modeling the background is to define the DE density and pressure as follows~\cite{Frusciante:2020xia}:
\begin{align}\label{eq:effective.density.pressure}
    \Tilde{P}_{\rm de}    &\equiv \Lambda + M_{p}^{2} \ddot{\Omega} + 2 M_{p}^{2} H \dot{\Omega} \nonumber \\
    \Tilde{\rho}_{\rm de} &\equiv 2c - \Lambda - 3M_{p}^{2} H \dot{\Omega}
\end{align}
This allows to change the parametrization eliminating $\Lambda(t)$ and $c(t)$ in favour of $\Tilde{\rho}_{\rm de}$ and $\Tilde{P}_{\rm de}$ so that Friedmann equations read:
\begin{align}
    3 M_{p}^{2} \Omega \left[ H^{2} + \frac{K}{a^{2}} \right] &= \Tilde{\rho}_{\rm de} + \rho_{\rm m} \nonumber \\
    M_{p}^{2} \Omega \left[ 2\dot{H}+3H^{2}+\frac{K}{a^{2}} \right] &= -\Tilde{P}_{\rm de} - P_{\rm m} 
\end{align}
and are thus formally analogous to the standard Friedmann Equations of $\Lambda CDM$.\\
The main issue with this definition is that the dark energy pressure and density do not satisfy the usual continuity equation, but instead obey the following one:
\begin{align}\label{eq:no_continuity}
    \Dot{\Tilde{\rho}}_{\rm de} + 3H(\Tilde{\rho}_{\rm de} + \Tilde{P}_{\rm de})= 3 M_{p}^{2}\Dot{\Omega} \left( H^2+\frac{K}{a^2} \right)
\end{align}
This implies that, even when $\Tilde{\rho}_{\rm de} = 0$, one still finds $\Tilde{P}_{\rm de} \neq 0$; that is, a non-zero dark energy pressure remains despite the absence of dark energy.\\
We therefore aim to introduce a new definition of dark energy pressure and density that satisfies the standard continuity equation for a perfect fluid $ \dot{\rho}_{\rm de}+3H(\rho_{\rm de}+P_{\rm de})=0 $, ensuring that when dark energy density vanishes, its pressure does as well.\\
In order to accomplish this, let us define the new dark energy pressure and density as:
\begin{align}\label{eq:new_definition}
    P_{\rm de}&\equiv \Tilde{P}_{\rm de}-M_p^2\left[(1+3\gamma)H\dot{\Omega}+\frac{\dot{\Omega}}{H}\frac{K}{a^2}+\gamma\ddot{\Omega}+\gamma \frac{\dot{\Omega}}{H}\dot{H}\right] \nonumber \\
    \rho_{\rm de}&\equiv \Tilde{\rho}_{\rm de} +3\gamma M_p^2H\dot{\Omega}
\end{align}
Where $\gamma$ is a free constant. In terms of EFT functions Eq.~\eqref{eq:new_definition} reads:
\begin{align}\label{rho_DE}
     &P_{\rm de}\equiv \Lambda +M_p^2\left[(1-\gamma)\ddot{\Omega}+(1-3\gamma)H \dot{\Omega}- \frac{\dot{\Omega}}{H}\frac{K}{a^2}-\gamma \frac{\dot{\Omega}}{H}\dot{H}\right]\nonumber\\ 
    &\rho_{\rm de}\equiv 2c-\Lambda-3M_{p}^{2}(1-\gamma)H\dot{\Omega}
\end{align}
With this new definition, $ P_{\rm de}$ and $\rho_{\rm de}$ satisfy the continuity equation and the modified Friedmann equations read:
\begin{align}\label{eq:Friedmann_new}
    3M_{p}^{2}\Omega\left[ H^{2}+\frac{K}{a^{2}}+\gamma \frac{\dot{\Omega}}{\Omega}H\right]&=\rho_{\rm de}+\rho_{\rm m}\nonumber\\
    M_p^2\Omega\left[2\dot{H}+3H^2+\frac{K}{a^2}+\gamma  \frac{\ddot{\Omega}}{\Omega}+(1+3\gamma)\frac{\dot{\Omega}}{\Omega}H+\gamma \frac{\dot{\Omega}}{\Omega}\frac{\dot{H}}{H}\right]&+M_p^2\frac{\dot{\Omega}}{H}\frac{K}{a^2}=-P_{\rm de}-P_{\rm m}  
\end{align} 
We refer to the set \{$H$, $\Omega$, $P_{\rm de}$, $\rho_{\rm de}$\} as the \emph{continuity-equation-compatible} (CEC) basis.\\
Note that $P_{\rm de}$ and $\rho_{\rm de}$ are not independent, but rather related by the standard continuity equation and can be both fixed by specifying $w_{\rm de}$.
Finally, let us verify that the definition of Eqs.~\eqref{rho_DE} has the correct $\Lambda CDM$ limit. Therefore, let us set $\Lambda(t)=\Lambda$, $c(t)=0$ and $\Omega(t)=1$ in Eq.~\eqref{rho_DE}:
\begin{align}
    \rho_{\rm de}&=\Lambda \nonumber\hspace{2cm}P_{\rm de}=-\Lambda
\end{align}
That corresponds to the standard $\Lambda$CDM equation of state for dark energy, $\rho_{\rm de}=-P_{\rm de}$.\\[0.1cm]

%%%%%%%%%%%%%%%%%%%%%%%%%%%%%%%%%%%%%%%%%%%%
\section{Numerical Implementation}
%%%%%%%%%%%%%%%%%%%%%%%%%%%%%%%%%%%%%%%%%%%%

To test our theoretical model, we implemented it using EFTCAMB~\cite{Hu:2016xia}, a powerful and versatile tool capable of evolving the full dynamics of linear scalar perturbations for a wide range of single-field dark energy and modified gravity models, once the model is mapped into the EFT formalism.\\
    To match the code's notation, we followed two main steps. First, we switched from cosmic time $t$ to conformal time $\tau$, so, from now on, dotted derivatives represent derivatives with respect to $\tau$, while primed derivatives refer to derivatives with respect to $a$. Second, we replaced $\Omega$ with $1 + \Omega$ to avoid setting the coefficient in front of the Ricci scalar to zero. Alongside this change of variables, some quantities such as the Hubble function must also be redefined: $\mathcal{H}=\frac{\dot a}{a}$ is now the Hubble function in conformal time, and $\dot{\mathcal{H}}$ its conformal time derivative.\\
With this clarification in place, to complete the implementation of the model we now need to determine the expressions for $\mathcal{H}^2$, $\dot{\mathcal{H}}$, $c$, and $\Lambda$, given $\rho_{\rm m}$, $P_{\rm m}$, $\rho_{\rm DE}$, $P_{\rm DE}$, $\gamma$, $K$, and $\Omega$.\\
The result is:

\begin{align}\label{eq:H2.conformal}
    \mathcal{H}^2\,=\,
    \frac{1}{1+\Omega+a\gamma\Omega'}\Bigl[a^2\frac{\rho_{\rm de}+\rho_{m}}{3M_P ^2}-\left(1+\Omega\right)K\Bigr]
\end{align}

\begin{align}\label{eq:Hdot.conformal}
    \dot {\mathcal{H}}\,=\,
    -\frac{1}{2(1+\Omega+a\gamma\Omega')}&\Bigl[a^2\frac{P_m+P_{\rm de}}{M_P ^2}+\mathcal{H}^2\Bigl(1+\Omega
    +(1+2\gamma)\Omega'+a^2\gamma\Omega''\Bigr)\nonumber\\
    &+(1+\Omega+a\Omega')K\Bigr]
\end{align}

\begin{align}\label{eq:c.conf}
    c\,=\,\frac{\rho_{\rm de}+P_{\rm de}}{2}-\frac{M^2_p}{2a}&\left[ \mathcal{H}^2\Bigl((1-\gamma)a\Omega''-(2-\gamma)\Omega'\Bigr)+\dot{\mathcal{H}}(1-2\gamma)- K\Omega' \right]
\end{align}

\begin{align}\label{eq:Lambda.conf}
    \Lambda\,=\,\frac{\rho_{\rm de}-P_{\rm de}}{2}+\frac{M^2_p}{2a}&\left[ \mathcal{H}^2\Bigl((1-\gamma)a\Omega''+(4-5\gamma)\Omega'\Bigr)+\dot{\mathcal{H}}(1-2\gamma)- K\Omega' \right]
\end{align}

\begin{comment}
    From Eqs. \eqref{eq:Friedmann_new} is straight to gain $H^2$ and $\dot{H}$: \textcolor{red}{Fede: se non ho sbagliato $\Omega'=\frac{\dot{\Omega}}{H}$ e $\Omega''=\frac{\ddot{\Omega}}{H^2}-\frac{\dot{H}\dot{\Omega}}{H^3}$}
\begin{align}
    H^2&=\frac{\rho_{\rm de}+\rho_{\rm m}}{3M_p^2\Omega}-\gamma \frac{\Omega '}{\Omega}H^2-\frac{k}{a^2} \textcolor{red}{\;Fede: cambio \, di \, base \, fatto \, fino \, a \, qui}\\
    \dot{H}&=-\frac{\frac{P_{\rm de}+P_{\rm m}}{2M_p^2}+\frac{3+3\Omega +\Omega ' }{2}H^2+\frac{1+\Omega+\Omega ' }{2}\frac{K}{a^2}}{1+\Omega }       
\end{align}
Where $\Omega '=\frac{d \Omega}{d \ln{a}}$. Finally, by adding and subtracting Eqs.~\eqref{rho_DE}, one obtains $c$ and $\Lambda$:
\begin{align}
    c &=-\frac{M_p^2}{2}\left[-2H^2\Omega '+\dot{H}\Omega'+H^2 \Omega ''-\frac{K}{a^2}\Omega '\right]+\frac{\rho_{\rm de}+P_{\rm de}}{2} \\
    \Lambda &=+\frac{M_p^2}{2}\left[4H^2\Omega'+\dot{H}\Omega '+H^2 \Omega ''-\frac{K}{a^2}\Omega '\right]+\frac{\rho_{\rm de}-P_{\rm de}}{2}
\end{align}
\end{comment}

%%%%%%%%%%%%%%%%%%%%%%%%%%%%%%%%%%%%%%%%%%%%
\section{Numerical Results}

 To test the implementation, we decided to plot the comparison between the Hubble function of the EFT of DE model, and the $\Lambda$CDM one, for this choice of the parameters:
 \begin{align}
     &\Omega\,=\,[-0.1\,,\,-0.05\,,\,0\,,\,0.05\,,\,0.1]\nonumber\\
     &\omega_{\rm de}\,=\,-1\nonumber\\
     &\gamma\,=\,0\nonumber\\
     &K\,=\,0
 \end{align}
 What we obtained is:
 \begin{figure}[h!]
     \centering
     \includegraphics[width=1\linewidth]{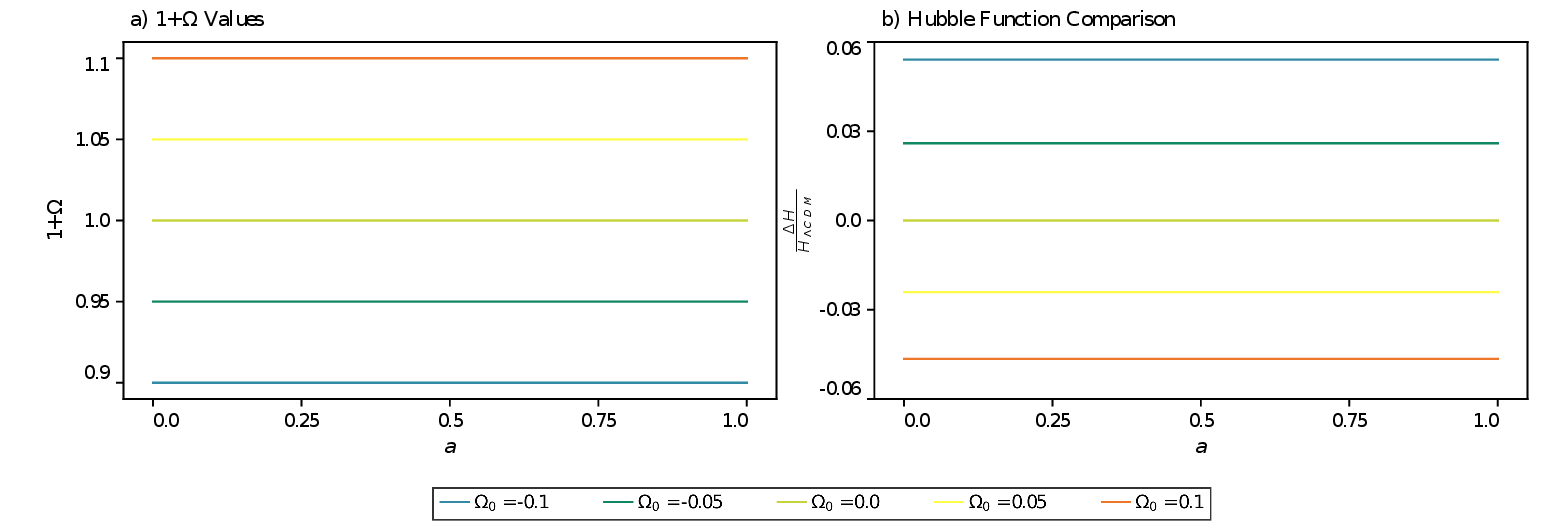}
     \caption{a) Different $\Omega$ constant values. b) Comparison, for each $\Omega$ value, between the EFT of DE Hubble function and the $\Lambda$CDM one; to be more specific, we computed $\frac{\mathcal{H}_{\rm de}- \mathcal{H}_{\rm \Lambda CDM}}{\mathcal{H}_{\rm \Lambda CDM}}$.}
     \label{fig:hubble.comparison}
 \end{figure}
 
As we can see, changing the value of $\Omega$, keeping the density of Dark Energy constant ($\omega_{\rm de}=-1$), makes the Hubble function shift with respect to the $\Lambda$CDM values, as expected. Moreover, the further the value of $\Omega$ is from $0$, the larger is the shift of $\mathcal{H}$ with respect to $\mathcal{H}_{\rm \Lambda CDM}$.

%%%%%%%%%%%%%%%%%%%%%%%%%%%%%%%%%%%%%%%%%%%%

\section{Conclusion}
In this work, we have addressed the ambiguity inherent in the choice of basis for the Effective Field Theory of Dark Energy. Starting from the standard EFT background action, we examined the standard parametrisation and highlighted its physical limitations. Then, we proposed a new definition of the dark energy density and pressure that satisfies the standard continuity equation, thus defining the \emph{continuity-equation-compatible} (CEC) basis. This redefinition allows for a more transparent physical interpretation by replacing two of the three background EFT functions, $c(t)$ and $\Lambda(t)$, with quantities that carry direct physical meaning.
\\
In addition, we outlined the numerical implementation of our framework, providing explicit expressions to recover the full EFT background dynamics from a physically motivated set of input functions.\\
Altogether, our results contribute to a more physically grounded formulation of the EFT of Dark Energy, with the potential to improve both theoretical understanding and the interpretation of cosmological observations within this framework.

\acknowledgments
\noindent
M.R. acknowledges financial support from the INFN InDark initiative.\\
E.M acknowledges that part of the analysis has been performed on the parallel computing cluster of the Open Physics Hub (https://site.unibo.it/openphysicshub/en) at the Physics and Astronomy Department in Bologna.

\end{document}